\documentclass[pra,a4paper,aps,onecolumn,showpacs,superscriptaddress,groupedaddress]{revtex4}
\usepackage{ae} 
\usepackage[T1]{fontenc}
\usepackage[ansinew]{inputenc}
\usepackage{amsmath}
\usepackage{amssymb}
\usepackage{graphicx}
\usepackage{color}
\usepackage[colorlinks]{hyperref}
\usepackage{lscape}

\begin{document}
\title{Character of superposed states under deterministic LOCC}
\author{Amit Bhar}\email{bhar.amit@yahoo.com}
\affiliation{Department of Mathematics, Jogesh Chandra Chaudhuri College, 30, Prince Anwar Shah Road, Kolkata-700033, India.}
\author{Ajoy Sen}\email{ajoy.sn@gmail.com}
\affiliation{Department of Applied Mathematics, University of Calcutta, 92, A.P.C. Road, Kolkata-700009, India.}
\author{Debasis Sarkar}
\email{dsappmath@caluniv.ac.in}
\affiliation{Department of Applied Mathematics, University of Calcutta, 92, A.P.C. Road, Kolkata-700009, India.}

\begin{abstract}
In this paper we investigate the effect of superposition of states on local conversion of pure bipartite states under deterministic LOCC. We are able to form a bridge between comparable and incomparable classes of states through the linear superposition of states. For example, if we consider two pairs of incomparable states, then their superposition may result into a comparable pair of states. We investigate many such cases and provide some of the results in tabular form. We also investigate the entanglement behavior of such classes of states, specifically their monotone nature. Finally we provide some  bounds of different measures of entanglement based on the idea of comparability and incomparability under deterministic LOCC.
\end{abstract}

\date{\today}
\pacs{03.67.Hk, 03.65.Ud, 03.65.Ta. \\ Keywords: LOCC, Entanglement, Superposition, Incomparability.}

\maketitle

\section{Introduction}
Quantum entanglement is one of the most puzzling, useful yet experimentally verified feature of quantum states. This was first noticed by Einstein, Podolsky and Rosen (EPR) \cite{rosen} and then $Sch\ddot{o}dinger$\cite{sch} coined the name \textit{`Entanglement"} for this phenomena. Later, Bell \cite{bell} observed that entangled state could be used to show the violation of a type of inequality which every local physical theory should obey. Quantum entanglement is also useful for performing many informational and computational tasks like Teleportation, Dense Coding etc.,\cite{zurek,nielsen,tele} which are otherwise impossible. Now, to understand behavior of quantum entanglement better, we need to probe different aspects of entanglement properly\cite{horo,Enk}.
Physicists have tried to observe the underline physics of quantum entanglement\cite{brus,linear} and suggested many algorithms and concepts to prove some new results.

In the paper \cite{super}, Linden \textit{et. al.} have raised the following problem: \textit{Suppose a bipartite quantum state} $|\Gamma\rangle$ \textit{and a certain decomposition of it as a superposition of two other states is given. In} $|\Gamma\rangle=\alpha|\psi\rangle +\beta|\phi\rangle$ \textit{what is the relation between the entanglement of} $|\Gamma\rangle$ \textit{and those of the two constituent states in the superposition}? They also considered the following two examples to illustrate the above problem. One is $|\gamma\rangle=\frac{1}{\sqrt{2}}|00\rangle+\frac{1}{\sqrt{2}}|11\rangle$ and the other is $|\gamma'\rangle = \frac{1}{\sqrt{2}}|\phi^{+}\rangle+\frac{1}{\sqrt{2}}|\phi^{-}\rangle$ where $|\phi^{\pm}\rangle=\frac{1}{\sqrt{2}}|00\rangle\pm\frac{1}{\sqrt{2}}|11\rangle$ are two common Bell states. The first one i.e., $|\gamma\rangle$ is a maximally entangled state but each constituent state is fully separable\cite{unitary,gisin}. That is, superposition of fully separable states can form a maximally entangled state and the second example shows exactly the opposite to that of the first, where $|\gamma'\rangle$ is separable but each constituent state is maximally entangled. Therefore, through the superposition of states one can find new physical insights regarding entanglement behaviour of differently correlated states. One can also ask what is the effect of superposed states of different kinds under LOCC, in particular, under deterministic LOCC. Is there any typical feature of superposed states under LOCC? In this work we want to study the effect of superposition on states that are deterministically convertible to another state under LOCC and also the states which are not convertible. Quite interestingly we find one can generate pair of comparable states\cite{flip,incompare,incomfi} through the superposition of incomparable states depending upon some specific conditions. The behavior of entanglement also changes with specific conditions under such superpositions of different kind of comparable or incomparable states. We have explicitly calculated in $3 \times 3$ system five different kinds of superpositions of comparable or incomparable states and used concurrence as the measure of entanglement to compare their entanglement behavior.\\

In \cite{super}, Linden \textit{et. al.} employed von-Neumann entropy of the reduced system as the entanglement measure($E$) and for this measure they found some upper bounds  of the entanglement of the superposed state in terms of the states being superposed. Yu \textit{et. al.}\cite{con1} have studied the concurrence of superposition and presented both upper bound and lower bound on the concurrence of superposition. Ou \textit{et. al.}\cite{negativity} gave an upper bound on the negativity of superposition. Niset and Cerf \cite{con} showed the lower and upper bounds in a simpler form. D.Cavalvanti \textit{et. al.} \cite{multi}, Song \textit{et. al.} \cite{multi1} and Yu \textit{et. al.}\cite{bound1} have investigated the entanglement of superpositions for multipartite quantum states by employing different entanglement measures. Gour \cite{ent} reconsidered the question in \cite{super} and presented tighter upper and lower bounds. Here we will show some new bounds on some of the other entanglement measures like, Negativity($N$), Logarithmic Negativity($LN$), Reyni-Entropy($S_{\delta}$). \\

This paper is organized as follows: in section \textbf{II}, we will discuss about some useful notions regarding superposition of states and entanglement. In section \textbf{III}, we will discuss the concept of incomparability. Section \textbf{IV} and \textbf{V} are devoted to discuss the main results and some nice illustrations on the bounds of different measures of entanglement. The paper is ended with a brief conclusion of our results.

\section{Superposition of States and Concurrence}

Quantum mechanics is inherently a linear theory and superposition is deeply related to this linear structure of quantum systems. Entanglement is a manifestation of quantum superposition whenever one deals with the composite systems. Superposition of two pure product state may give rise to an entangled state and quite contrary to it one can get pure product states from the superposition of entangled states only. It is clear that if someone tries to explain superposition of states as a physical process then it should not be local, as entanglement may be created or increased in this process.
Now, for pure bipartite states, apart from the entropy of entanglement there is an useful quantifier of entanglement which is called generalized concurrence($C$). For a separable state it is zero. For a two-qubit state $\rho_{AB}$ it is calculated by $C(\rho_{AB})=\max \{\lambda_{1}-\lambda_{2}-\lambda_{3}-\lambda_{4},0\}$ where $\lambda_{i}$'s, $i=1,2,3,4$ are the square root of the eigenvalues of $\rho\tilde{\rho}$ in decreasing order where $\tilde{\rho}=\sigma_{y}\otimes\sigma_{y}\rho_{AB}^{*}\sigma_{y}\otimes\sigma_{y}$ and $*$ denotes conjugate operation. For higher order systems generalized concurrence is defined by,

\begin{equation}
C(\rho_{AB})= \sqrt{2(1-Tr{{\rho_A}^2})}
\end{equation}
where $\rho_A$ is the reduced density matrix of obtained by tracing out the subsystem $B$. For a pure bipartite state $|\xi\rangle_{AB}$  of $d_1 \times d_2$  system with Schmidt form
$|\xi\rangle_{AB}=\sum_i^{\min\{d_1,d_2\}} \sqrt{\mu_i}|i\rangle_{A} |i\rangle_{B}$, where $\{\mu_i ~;~~i=1,2,\cdots \}$ are non-negative Schmidt coefficients and $\{|i\rangle_A\}$, $\{ |i\rangle_B\}$ being the orthonormal bases for subsystems $A$ and $B$ respectively, the generalized
concurrence $C(|\xi\rangle_{AB})$  turns out to be,
\begin{equation}
C^{2}(|\xi\rangle_{AB})=4\sum_{i<j} \mu_i \mu_j= 2(1-\sum_{i=1}^{\min\{d_1,d_2\}}
\mu_i^{2})
\end{equation}
varies smoothly from $0$ for pure product states to $2\frac{d-1}{d}$ for maximally entangled pure bipartite states of Schmidt rank $d$.

\section{Notion of Incomparability}

Now, before going to present our results, we first mention the condition for a pair of pure bipartite states to be incomparable with each other. The notion of incomparability of a pair of bipartite pure states is a consequence of Nielsen's \cite{nielsen,nielsen1} famous \textit{majorization criterion}. To illustrate it, we consider the deterministic local conversion of the pure bipartite state $|\chi\rangle$ to $|\eta\rangle$ shared between two parties, say, Alice and Bob. We write the pair $(|\chi\rangle$,$|\eta\rangle)$ in their Schmidt bases $\{|i_A\rangle ,|i_B\rangle\}$ with decreasing order of Schmidt coefficients: $|\chi\rangle=\sum_{i=1}^{d} \sqrt{\gamma_{i}} |i_A i_B\rangle$, $|\eta\rangle=\sum_{i=1}^{d} \sqrt{\delta_{i}} |i_A i_B\rangle.$ The Schmidt vectors corresponding to the states $|\chi\rangle$ and $|\eta\rangle$ are $\lambda_\chi\equiv(\gamma_1,\gamma_2,\cdots,\gamma_d)$ and $\lambda_\eta\equiv(\delta_1,\delta_2,\cdots,\delta_d)$. From Nielsen's criterion, $|\chi\rangle\rightarrow| \eta\rangle$ is possible with certainty under LOCC if and only if $\lambda_\chi$ is majorized by $\lambda_\eta,$ (denoted by
$\lambda_\chi\prec\lambda_\eta$), i.e.,
\begin{equation}
\begin{array}{lcl}\sum_{i=1}^{k}\gamma_{i}\leq
\sum_{i=1}^{k}\delta_{i}~ ~\forall~ ~k=1,2,\cdots,d
\end{array}
\end{equation}

The above result has a direct consequence in the entanglement behavior of the states involved. If $|\chi\rangle\rightarrow |\eta\rangle$ is possible under deterministic LOCC, then $E(|\chi\rangle)\geq E(|\eta\rangle)$ [where $E(\cdot)$ is the entropy of entanglement]. Now in case of failure of the above criterion, we denote it by $|\chi\rangle\not\rightarrow |\eta\rangle$. But it may happen
that $|\eta\rangle\rightarrow |\chi\rangle$ is possible under deterministic LOCC. If both $|\chi\rangle\not\rightarrow |\eta\rangle$ and
$|\eta\rangle\not\rightarrow |\chi\rangle$ do not hold, we denote it by $|\chi\rangle\not\leftrightarrow |\eta\rangle$ and call $(|\chi\rangle, |\eta\rangle)$ as a pair of incomparable states. The existence of incomparable pair of states starts from $3\times 3$ systems. For our purpose, we require explicitly the criterion of incomparability for a pair of pure bipartite states $|\chi\rangle, |\eta\rangle$ of $3\times 3$ system. Suppose the Schmidt vectors corresponding to the two states are $(\gamma_1, \gamma_2, \gamma_3)$ and $(\delta_1, \delta_2, \delta_3)$ respectively, where $\gamma_1> \gamma_2> \gamma_3>0~,~\delta_1> \delta_2> \delta_3>0~,~\gamma_1+ \gamma_2+ \gamma_3=1=\delta_1+ \delta_2+ \delta_3$. Then $|\chi\rangle, |\eta\rangle$ are incomparable whenever\cite{som} either of the following relations hold.

\begin{equation}
\begin{array}{lcl}
(i) ~~\gamma_1 > \delta_1 ~~\text{and}~~ \gamma_3 > \delta_3\\
(ii)~~ \delta_1 > \gamma_1 ~~\text{and}~~ \delta_3 > \gamma_3
\end{array}
\end{equation}

\section{Main Results}

Consider the states shared between two parties say, A and B,
\begin{equation}\label{}
|\Gamma\rangle_{AB}=\alpha|\psi\rangle_{AB} +\beta|\phi\rangle_{AB}
\end{equation}
where $\alpha^{2}+\beta^{2}=1$ and $\alpha,\beta$ are non-negative real number and also consider the state

\begin{equation}\label{}
|\Gamma'\rangle_{AB}=\alpha'|\psi'\rangle_{AB} +\beta'|\phi'\rangle_{AB}
\end{equation}
where $\alpha'^{2}+\beta'^{2}=1$ with non-negative real $\alpha',\beta'$ and further assume that $\langle\psi|\phi\rangle_{AB}=0$ ; $\langle\psi'|\phi'\rangle_{AB}=0$.  Explicitly, suppose $|\psi\rangle_{AB}$, $|\psi'\rangle_{AB}$, $|\phi\rangle_{AB}$ and $|\phi'\rangle_{AB}$ may be expressed as follows:
\begin{eqnarray}
|\psi\rangle_{AB}&=&\sum_{i=0}^{2}\sqrt{a_{i}}|ii\rangle_{AB} \\
|\phi\rangle_{AB}&=&\sum_{j=0}^{2}\sqrt{b_{j}}|jj\rangle_{AB}\\
|\psi'\rangle_{AB}&=&\sum_{i=0}^{2}\sqrt{\alpha_{i}}|ii\rangle_{AB}\\
|\phi'\rangle_{AB}&=&\sum_{j=0}^{2}\sqrt{\beta_{j}}|jj\rangle_{AB}
\end{eqnarray}
We will now discuss the entanglement behavior of the superposed states imposing some restrictions on $\alpha,\beta,\alpha',\beta',$ and also on $a_{i},b_{i},\alpha_{i}, \beta_{i}$ for all i=0,1,2, case by case.\\

\textbf{\\\underline{CASE:I}\\ }

Consider that the states  ($|\psi\rangle_{AB}$, $|\psi'\rangle_{AB}$) and ($|\phi\rangle_{AB}$ ,$|\phi'\rangle_{AB}$) are mutually incomparable to each other under deterministic LOCC. We test whether this incomparability of states become the global phenomenon of $|\Gamma\rangle_{AB}$ and $|\Gamma'\rangle_{AB}$ or not, i.e., the states $|\Gamma\rangle_{AB}$ and $|\Gamma'\rangle_{AB}$ are incomparable under LOCC to each other or not. We illustrate all the results through the following tabular form:\\\\

\begin{center}
\textbf{Table-1}
\begin{center}
\textbf{RELATIONSHIP BETWEEN $|\Gamma\rangle_{AB}$ AND $|\Gamma'\rangle_{AB}$ AS PROVIDED IN CASE:I }
\end{center}
\end{center}

\begin{center}
\begin{tabular}{|c|c|c|}
  \hline
  \textbf{RESTRICTIONS ON}  & \textbf{SOME OTHER} & \textbf{NATURE OF THE PAIR} \\ $\alpha,\beta$ & \textbf{CONSIDERATIONS} & ($|\Gamma \rangle_{AB}$ and $  |\Gamma'\rangle_{AB}$) \\
  \hline

  $\alpha =\alpha' \text{ and } \beta=\beta'$ & -- & INCOMPARABLE  \\
  \hline

  $\alpha >\alpha' \text{ and } \beta<\beta'$& $(\beta^{2}b_{0}>\beta'^{2}\beta_{0})$ \text{ and }
  $(\beta^{2}b_{2}>\beta'^{2}\beta_{2})$&INCOMPARABLE \\
  \hline

  $\alpha <\alpha' \text{and } \beta>\beta'$ &  $(\alpha^{2}a_{0}>\alpha'^{2}\alpha_{0})\text{ and } (\alpha^{2}a_{2}>\alpha'^{2}\alpha_{2})$&INCOMPARABLE \\
  \hline

  $\alpha >\alpha' \text{ and } \beta<\beta'$& $(\beta^{2}b_{0}>\beta'^{2}\beta_{0})$;
  $(\beta^{2}b_{2}>\alpha'^{2}\alpha_{2})$ \text{ and }
  $(\alpha^{2}a_{2}>\beta'^{2}\beta_{2})$& COMPARABLE \\ \hline
  $\alpha <\alpha' \text{ and } \beta>\beta'$ &  $(\alpha^{2}a_{0}>\alpha'^{2}\alpha_{0})$;$(\alpha^{2}a_{2}>\beta'^{2}\beta_{2})$
  \text{ and } $(\beta^{2}b_{2}>\alpha'^{2}\alpha_{2})$&COMPARABLE \\

  \hline

\end{tabular}\\
\end{center}

i.e., in some cases the pair $|\Gamma\rangle_{AB}$ and $|\Gamma'\rangle_{AB}$ remain incomparable under deterministic LOCC. However, for some cases we find the pair as comparable. The corresponding entanglement behavior are as follows;

\begin{center}
\textbf{Table-1A}
\end{center}
\begin{center}
\textbf{RELATIONSHIP OF ENTANGLEMENT BETWEEN $|\Gamma\rangle_{AB}$ AND $|\Gamma'\rangle_{AB}$}
\end{center}

\begin{center}
\begin{tabular}{|c|c|c|c|}

\hline
 \textbf{ RESTRICTIONS}  &\textbf{NATURE OF THE PAIR} & \textbf{SOME OTHER} & \textbf{NATURE OF THE PAIR}\\\textbf{ON} $\alpha,\beta$ & $(|\Gamma \rangle_{AB}$ and $| \Gamma'\rangle_{AB}$)& \textbf{ CONSIDERATIONS}&($C^{2}(|\Gamma\rangle_{AB})$ and $C^{2}(|\Gamma'\rangle_{AB}$) \\
 \hline

  $\alpha =\alpha' \text{ and } \beta=\beta'$ &INCOMPARABLE  & $(\alpha'\sqrt{\alpha_{1}}+\beta'\sqrt{\beta_{1}})^{2}>\frac{1}{2}$ &$C^{2}(|\Gamma\rangle_{AB}) > C^{2}(|\Gamma'\rangle_{AB})$ \\
  \hline

  $\alpha >\alpha' \text{ and } \beta<\beta'$&INCOMPARABLE&$\alpha^{2}a_{1}<\alpha'^{2}\alpha_{1}$& $C^{2}(|\Gamma\rangle_{AB}) > C^{2}(|\Gamma'\rangle_{AB})$\\&&$(\alpha'\sqrt{\alpha_{1}}+\beta'\sqrt{\beta_{1}})^{2}>\frac{1}{2}$&\\
  \hline

  $\alpha <\alpha' \text{ and } \beta>\beta'$&INCOMPARABLE&$\beta^{2}b_{1}<\beta'^{2}\beta_{1}$& $C^{2}(|\Gamma\rangle_{AB}) > C^{2}(|\Gamma'\rangle_{AB})$\\&&$(\alpha'\sqrt{\alpha_{1}}+\beta'\sqrt{\beta_{1}})^{2}>\frac{1}{2}$&   \\
  \hline

  $\alpha >\alpha' \text{ and } \beta<\beta'$&COMPARABLE&$\alpha^{2}a_{1}<\alpha'^{2}\alpha_{1}$& $C^{2}(|\Gamma\rangle_{AB}) > C^{2}(|\Gamma'\rangle_{AB})$\\&&$(\alpha'\sqrt{\alpha_{1}}+\beta'\sqrt{\beta_{1}})^{2}\lessgtr\frac{1}{2}$&\\
  \hline

  $\alpha <\alpha' \text{ and } \beta>\beta'$&COMPARABLE&$\beta^{2}b_{1}<\beta'^{2}\beta_{1}$& $C^{2}(|\Gamma\rangle_{AB}) > C^{2}(|\Gamma'\rangle_{AB})$\\&&$(\alpha'\sqrt{\alpha_{1}}+\beta'\sqrt{\beta_{1}})^{2}\lessgtr\frac{1}{2}$&\\

  \hline
\end{tabular}\\
\end{center}

i.e., in all the above subcases generalized concurrence of $|\Gamma\rangle_{AB}$ is greater than that of $|\Gamma'\rangle_{AB}$.

\textbf{\\\underline{CASE:II}\\ }

Next we consider the states  $(|\psi\rangle_{AB}$, $|\psi'\rangle_{AB})$ as a comparable pair and $(|\phi\rangle_{AB}$, $ |\phi'\rangle_{AB})$ as mutually incomparable to each other under deterministic LOCC. The behavior of $|\Gamma\rangle_{AB}$ and $|\Gamma'\rangle_{AB}$ is illustrated in the following tabular form:

\begin{center}
\textbf{Table-2}
\end{center}
\begin{center}
\textbf{RELATIONSHIP BETWEEN $|\Gamma\rangle_{AB}$ AND $|\Gamma'\rangle_{AB}$ }
\end{center}

\begin{center}
\begin{tabular}{|c|c|c|}
 \hline
 \textbf{RESTRICTIONS ON } & \textbf{SOME OTHER} & \textbf{NATURE OF THE PAIR} \\
 $\alpha,\beta$& \textbf{CONSIDERATIONS}&($|\Gamma\rangle_{AB}$ and $|\Gamma'\rangle_{AB}$)\\
 \hline
 $\alpha =\alpha' \text{ and } \beta=\beta'$ &$(a_{2}b_{2}>\alpha_{2}\beta_{2})$  & INCOMPARABLE  \\
 \hline

 $\alpha >\alpha' \text{ and } \beta<\beta'$& $(\beta^{2}b_{0}>\beta'^{2}\beta_{0})$; $(\alpha^{2}a_{2}>\alpha'^{2}\alpha_{2})$\text{ and }
 $(\beta^{2}b_{2}>\beta'^{2}\beta_{2})$&INCOMPARABLE \\
 \hline

 $\alpha <\alpha' \text{ and } \beta>\beta'$& $(\alpha^{2}a_{0}>\beta'^{2}\beta_{0})$;
 $(\beta^{2}b_{0}>\alpha'^{2}\alpha_{0})$& \\& and
 $(\alpha^{2}a_{2}>\beta'^{2}\beta_{2}); (\beta^{2}b_{2}>\alpha'^{2}\alpha_{2})$ & INCOMPARABLE \\& OR
 $(\alpha^{2}a_{0}<\beta'^{2}\beta_{0})$;
 $ (\beta^{2}b_{0}<\alpha'^{2}\alpha_{0})$ & \\& and
 $(\alpha^{2}a_{2}<\beta'^{2}\beta_{2}); (\beta^{2}b_{2}<\alpha'^{2}\alpha_{2})$ &  \\
 \hline

 $\alpha =\alpha' \text{ and } \beta=\beta'$ &$(a_{2}b_{2}<\alpha_{2}\beta_{2})$  & COMPARABLE  \\
 \hline

 $\alpha >\alpha' \text{ and } \beta<\beta'$& $(\beta^{2}b_{0}>\beta'^{2}\beta_{0})$; $(\alpha^{2}a_{2}<\alpha'^{2}\alpha_{2})$\text{ and }
 $(\beta^{2}b_{2}<\beta'^{2}\beta_{2})$&COMPARABLE \\
 \hline

 $\alpha <\alpha' \text{ and } \beta>\beta'$& $(\alpha^{2}a_{0}>\beta'^{2}\beta_{0})$;
 $(\beta^{2}b_{0}>\alpha'^{2}\alpha_{0})$& \\&and
 $(\alpha^{2}a_{2}<\beta'^{2}\beta_{2}); (\beta^{2}b_{2}<\alpha'^{2}\alpha_{2})$ & COMPARABLE \\& OR
 $(\alpha^{2}a_{0}<\beta'^{2}\beta_{0})$;
 $(\beta^{2}b_{0}<\alpha'^{2}\alpha_{0})$ & \\& and
 $(\alpha^{2}a_{2}>\beta'^{2}\beta_{2}); (\beta^{2}b_{2}>\alpha'^{2}\alpha_{2})$ &  \\
 \hline

\end{tabular}\\
\end{center}

i.e., incomparability may be preserved in some subcases and one can also break the incomparability by superposing the states in some other subcases.
\pagebreak
\begin{center}
\textbf{Table-2A}
\end{center}
\begin{center}
\textbf{RELATIONSHIP OF ENTANGLEMENT BETWEEN
$|\Gamma\rangle_{AB}$ AND $|\Gamma'\rangle_{AB}$}
\end{center}

\begin{center}
\begin{tabular}{|c|c|c|c|}
 \hline
 \textbf{RESTRICTIONS}&\textbf{NATURE OF THE}&\textbf{SOME OTHER} &\textbf{NATURE OF THE PAIR}\\\textbf{ON} $\alpha,\beta$&\textbf{PAIR}($|\Gamma \rangle_{AB}$ and $ | \Gamma'\rangle_{AB}$)& \textbf{ CONSIDERATIONS}&($C^{2}(|\Gamma\rangle_{AB})$ and $C^{2}(|\Gamma'\rangle_{AB}$) \\

 \hline
 $\alpha =\alpha' \text{ and } \beta=\beta'$ &INCOMPARABLE  & $(\alpha'\sqrt{\alpha_{1}}+\beta'\sqrt{\beta_{1}})^{2}>\frac{1}{2}$ &$C^{2}(|\Gamma\rangle_{AB}) > C^{2}(|\Gamma'\rangle_{AB})$ \\
 \hline

 $\alpha >\alpha' \text{ and } \beta<\beta'$&INCOMPARABLE&$(\alpha^{2}a_{1}<\alpha'^{2}\alpha_{1})$ and & \\&& $(\alpha'\sqrt{\alpha_{1}}+\beta'\sqrt{\beta_{1}})^{2}>\frac{1}{2}$& $C^{2}(|\Gamma\rangle_{AB}) > C^{2}(|\Gamma'\rangle_{AB})$\\
 \hline

 $\alpha <\alpha' \text{ and } \beta>\beta'$&INCOMPARABLE&$(\beta^{2}b_{1}<\beta'^{2}\beta_{1})$ \text{ and }& \\&&
 $(\alpha'\sqrt{\alpha_{1}}+\beta'\sqrt{\beta_{1}})^{2}>\frac{1}{2}$& $C^{2}(|\Gamma\rangle_{AB}) > C^{2}(|\Gamma'\rangle_{AB})$\\
 &&\textbf{OR}&
 \\&&($\beta^{2}b_{1}>\beta'^{2}\beta_{1}$)\text{ and }& \\&& $(\alpha^{2}a_{1}>\alpha'^{2}\alpha_{1})$&\\&& $(\alpha\sqrt{a_{1}}+\beta\sqrt{b_{1}})^{2}>\frac{1}{2}$&
 \\&&\textbf{OR}& \\&&($\beta^{2}b_{1}<\beta'^{2}\beta_{1}$)\text{ and }& \\&& $\alpha^{2}a_{1}<\alpha'^{2}\alpha_{1}$ &\\&&$(\alpha'\sqrt{\alpha_{1}}+\beta'\sqrt{\beta_{1}})^{2}>\frac{1}{2}$&\\ \hline

 $\alpha =\alpha' \text{ and } \beta=\beta'$ &COMPARABLE  & $(\alpha'\sqrt{\alpha_{1}}+\beta'\sqrt{\beta_{1}})^{2}>\frac{1}{2}$ &$C^{2}(|\Gamma\rangle_{AB}) > C^{2}(|\Gamma'\rangle_{AB})$ \\
 && \textbf{OR}&\\&& $(\alpha'\sqrt{a_{1}}+\beta'\sqrt{b_{1}})^{2}<\frac{1}{2}$&\\
 \hline

 $\alpha >\alpha' \text{ and } \beta<\beta'$&COMPARABLE&$(\alpha^{2}a_{1}<\alpha'^{2}\alpha_{1}$)\text{ and }& \\&&
 $(\alpha'\sqrt{\alpha_{1}}+\beta'\sqrt{\beta_{1}})^{2}>\frac{1}{2}$ & $C^{2}(|\Gamma\rangle_{AB}) >C^{2}(|\Gamma'\rangle_{AB})$\\
 &&\textbf{OR} &\\&&($\alpha^{2}a_{1}<\alpha'^{2}\alpha_{1}$)\text{ and }  & \\&& $(\alpha\sqrt{a_{1}}+\beta\sqrt{b_{1}})^{2}<\frac{1}{2}$&\\
 \hline

 $\alpha <\alpha' \text{ and } \beta>\beta'$ &COMPARABLE&($\beta^{2}b_{1}<\beta'^{2}\beta_{1}$)\text{ and } & \\&& $(\alpha'\sqrt{\alpha_{1}}+\beta'\sqrt{\beta_{1}})^{2}>\frac{1}{2}$& $C^{2}(|\Gamma\rangle_{AB}) > C^{2}(|\Gamma'\rangle_{AB})$\\
 &&\textbf{OR}&\\
 &&$(\beta^{2}b_{1}>\beta'^{2}\beta_{1})$\text{ and } & \\&&$(\alpha^{2}a_{1}>\alpha'^{2}\alpha_{1})$&\\
 &&$(\alpha\sqrt{a_{1}}+\beta\sqrt{b_{1}})^{2}>\frac{1}{2}$&
 \\&&\textbf{OR}&
 \\&&($\beta^{2}b_{1}<\beta'^{2}\beta_{1}$)\text{ and }& \\&& ($\alpha^{2}a_{1}<\alpha'^{2}\alpha_{1}$)&\\
 &&$(\alpha'\sqrt{\alpha_{1}}+\beta'\sqrt{\beta_{1}})^{2}>\frac{1}{2}$&\\
 &&\textbf{OR}&\\
 &&($\beta^{2}b_{1}<\beta'^{2}\beta_{1}$)\text{ and }& \\&& $(\alpha\sqrt{a_{1}}+\beta\sqrt{b_{1}})^{2}<\frac{1}{2}$&\\
 \hline

\end{tabular}\\
\end{center}

Here, in all the above subcases generalized concurrence of $|\Gamma\rangle_{AB}$ is also greater than that of $|\Gamma'\rangle_{AB}$.\\

\textbf{\underline{CASE:III}\\ }

Next we consider two states $|\Gamma\rangle_{AB}$ and $|\Gamma''\rangle_{AB}$, where $|\Gamma''\rangle_{AB}=\alpha'|\psi'\rangle_{AB} +\beta'|\phi\rangle_{AB}$ with the assumption that the states  ($|\psi\rangle_{AB}$, $|\psi'\rangle_{AB}$) are mutually incomparable to each other. We provide whether $|\Gamma\rangle_{AB}$ and $|\Gamma''\rangle_{AB}$ are incomparable under LOCC to each other or not through the following tabular form:
\pagebreak
\begin{center}
\textbf{Table-3}
\end{center}
\begin{center}
\textbf{RELATIONSHIP BETWEEN $|\Gamma\rangle_{AB}$ AND $|\Gamma''\rangle_{AB}$ AS PROVIDED IN CASE:III }
\end{center}

\begin{center}
\begin{tabular}{|c|c|c|}
 \hline

 \textbf{RESTRICTIONS ON } & \textbf{SOME OTHER} & \textbf{NATURE OF THE PAIR} \\
$\alpha,\beta$&\textbf{CONSIDERATIONS}&($|\Gamma\rangle_{AB}$ and $|\Gamma''\rangle_{AB}$)\\
 \hline

 $\alpha =\alpha' \text{ and } \beta=\beta'$ & -- & INCOMPARABLE  \\
 \hline
 $\alpha <\alpha' \text{ and } \beta>\beta'$& $(\alpha^{2}a_{0}>\alpha'^{2}\alpha_{0})$\text{ and } $(\alpha^{2}a_{2}>\alpha'^{2}\alpha_{2})$
 &INCOMPARABLE \\
 \hline

 $\alpha >\alpha' \text{ and } \beta<\beta'$& $(\alpha^{2}a_{2}>\beta'^{2}b_{2})$;
 $(\beta^{2}b_{2}>\alpha'^{2}\alpha_{2})$& INCOMPARABLE \\
 \hline

 $\alpha <\alpha' \text{ and } \beta>\beta'$& $(\alpha^{2}a_{2}<\beta'^{2}\beta_{2})\text{ and } (\beta^{2}b_{2}<\alpha'^{2}\alpha_{2})$ &COMPARABLE \\
 \hline

 $\alpha >\alpha' \text{ and } \beta<\beta'$& $(\alpha^{2}a_{2}<\beta'^{2}b_{2})$;
 $(\beta^{2}b_{2}<\alpha'^{2}\alpha_{2})$& COMPARABLE \\

 \hline
\end{tabular}\\
\end{center}

 Here, we consider the pair $|\Gamma\rangle_{AB}$ and $|\Gamma''\rangle_{AB}$ in such a manner to clarify the motivation of our work. We observe that the superposed states are comparable under deterministic LOCC for some cases even when the state $|\phi\rangle_{AB}$ remains same for both the states $|\Gamma\rangle_{AB}$ and $|\Gamma''\rangle_{AB}$. Thus linear superposition plays a vital role in interchanging the status of superposed states. The corresponding entanglement behavior are as follows;

\begin{center}
\textbf{Table-3A}
\end{center}
\begin{center}
\textbf{RELATIONSHIP OF ENTANGLEMENT BETWEEN $|\Gamma\rangle_{AB}$ AND $|\Gamma''\rangle_{AB}$}
\end{center}

\begin{center}
\begin{tabular}{|c|c|c|c|}

 \hline

 \textbf{RESTRICTION}&\textbf{NATURE OF THE}&\textbf{SOME OTHER} &\textbf{NATURE OF THE PAIR }\\
  $\textbf{ ON } \alpha,\beta$ &PAIR($|\Gamma \rangle_{AB}$ and $| \Gamma''\rangle_{AB}$)&  \textbf{CONSIDERATIONS}&($C^{2}(|\Gamma\rangle_{AB})$ and $C^{2}(|\Gamma''\rangle_{AB}$) \\
 &  &&\\
 \hline
 $\alpha =\alpha' \text{ and } \beta=\beta'$ &INCOMPARABLE  & $(\alpha'\sqrt{\alpha_{1}}+\beta'\sqrt{b_{1}})^{2}>\frac{1}{2}$ &$C^{2}(|\Gamma\rangle_{AB}) > C^{2}(|\Gamma''\rangle_{AB})$ \\
 \hline

 $\alpha >\alpha' \text{ and } \beta<\beta'$&INCOMPARABLE&($\alpha^{2}a_{1}<\alpha'^{2}\alpha_{1}$)\text{ and } & \\&& $(\alpha'\sqrt{\alpha_{1}}+\beta'\sqrt{b_{1}})^{2}>\frac{1}{2}$& $C^{2}(|\Gamma\rangle_{AB}) > C^{2}(|\Gamma''\rangle_{AB})$\\
 &&\textbf{OR}&\\
 &&$(\alpha\sqrt{\alpha_{1}}+\beta'\sqrt{b_{1}})^{2}>\frac{1}{2}$&\\
 \hline

 $\alpha <\alpha' \text{ and } \beta>\beta'$&INCOMPARABLE&$(\alpha'\sqrt{\alpha_{1}}+\beta\sqrt{b_{1}})^{2}>\frac{1}{2}$& $C^{2}(|\Gamma\rangle_{AB}) > C^{2}(|\Gamma''\rangle_{AB})$\\
 &  &&\\
 \hline

 $\alpha >\alpha' \text{ and } \beta<\beta'$&COMPARABLE&$(\alpha^{2}a_{1}<\alpha'^{2}\alpha_{1})$\text{ and } & \\&& $(\alpha'\sqrt{\alpha_{1}}+\beta'\sqrt{b_{1}})^{2}>\frac{1}{2}$& $C^{2}(|\Gamma\rangle_{AB}) >C^{2}(|\Gamma''\rangle_{AB})$\\
 &&\textbf{OR}&\\
 &&($\alpha^{2}a_{1}<\alpha'^{2}\alpha_{1}$)\text{ and }& \\&& $(\alpha\sqrt{a_{1}}+\beta\sqrt{b_{1}})^{2}<\frac{1}{2}$& \\
 &&\textbf{OR}&\\
 &&$(\alpha\sqrt{\alpha_{1}}+\beta'\sqrt{b_{1}})^{2}>\frac{1}{2}$&\\
 &&\textbf{OR}&\\
 &&$(\alpha'\sqrt{a_{1}}+\beta\sqrt{b_{1}})^{2}<\frac{1}{2}$&\\
 \hline

 $\alpha <\alpha' \text{ and } \beta>\beta'$&COMPARABLE&$(\alpha'\sqrt{\alpha_{1}}+\beta\sqrt{b_{1}})^{2}>\frac{1}{2}$& $C^{2}(|\Gamma\rangle_{AB}) > C^{2}(|\Gamma''\rangle_{AB})$\\
 &&\textbf{OR}&\\
 &&$(\alpha\sqrt{a_{1}}+\beta'\sqrt{b_{1}})^{2}<\frac{1}{2}$&\\
 &  &&\\
 &  &&\\
 \hline

\end{tabular}\\
\end{center}

\textbf{\\\\\underline{CASE:IV}\\ }

Now we consider the states $|\Gamma\rangle_{AB}$ and $|\Gamma''\rangle_{AB}$ in such a manner that the states  ($|\psi\rangle_{AB}$, $|\psi'\rangle_{AB}$)  are comparable to each other under deterministic LOCC. We have found that the superposed states $|\Gamma\rangle_{AB}$ and $|\Gamma''\rangle_{AB}$ are incomparable and comparable  under deterministic LOCC when $\alpha <\alpha' , \beta >\beta'$ and $\alpha >\alpha' , \beta<\beta'$ respectively by imposing some other restrictions on the Schmidt coefficients like the previous tables. The superposed states are only comparable in the case $\alpha =\alpha' \text{ and } \beta=\beta'$ without imposing any restriction on the Schmidt coefficients. All the possible combinations of the Schmidt coefficients,the monotonic nature of concurrence is observed like in the previous tables.

\textbf{\\\underline{CASE:V}\\ }

Lastly, we consider the two states $|\Gamma\rangle_{AB}$ and $|\Gamma'\rangle_{AB}$, shared between two parties A and B,are such that the states  ($|\psi\rangle_{AB}$, $|\psi'\rangle_{AB}$) and ($|\phi\rangle_{AB}$ and $|\phi'\rangle_{AB}$) are comparable pairs under deterministic LOCC.
For all the possible combinations of the Schmidt coefficients with some other restrictions on them, the states $|\Gamma\rangle_{AB}$ and $|\Gamma'\rangle_{AB}$ are sometimes incomparable to each other under deterministic LOCC.\\

\section{OBSERVATIONS ON THE BOUNDS OF SUPERPOSED STATES}
In this section, our aim is to find some new bounds on entanglement for the superposed states $|\Gamma\rangle_{AB}$, and  $|\Gamma'\rangle_{AB}$ by using different entanglement measures like, Negativity($N$), Logarithmic Negativity($LN$), Reyni-Entropy($S_{\delta}$). We derive some tight bounds and also study the behavior of the bounds for the corresponding measure through the notion of incomparability under deterministic LOCC.\\

According to our assumption, the component states of the states $|\Gamma\rangle_{AB}$ and $|\Gamma'\rangle_{AB}$ are orthogonal, i.e., $\langle\psi|\phi\rangle_{AB}=0$. Taking  Negativity($N$) as our entanglement measure, we have found the following forms of upper and lower bounds of entanglement for the superposed state  $|\psi\rangle_{AB}$ in terms of the entanglement of the constituent states (i.e., $|\psi\rangle_{AB}$ and $|\phi\rangle_{AB}$) and also in terms of the Schmidt coefficients of the states. We provide all the results in theorem form, however their proofs are quite straight forward.\\

\textbf{Theorem 1}: \emph{$\alpha^{2}N(|\psi\rangle_{AB})+\beta^{2}N(|\phi\rangle_{AB})\leq N(|\Gamma\rangle_{AB})\leq \alpha^{2}N(|\psi\rangle_{AB})+\beta^{2}N(|\phi\rangle_{AB})+\alpha \beta$}.\\

\textbf{Theorem 2}: \emph{$\frac{1}{2}[9(\alpha+\beta)^{2}\{\min(\mu)^{2}\}-1]\leq N(|\Gamma\rangle_{AB})\leq \frac{1}{2}[9(\alpha+\beta)^{2}\{\max(\mu)^{2}\}-1]$} where $\min(\mu)$ and $\max(\mu)$ denote respectively the least and greatest of the numbers $\{\sqrt{a_{i}},\sqrt{b_{i}}\}_{i=0}^{2}$. \\

The proof of the above two theorems can be done easily by considering the definition of Negativity and using some basic results on inequality. Similar types of bounds can also be derived for the other two measures Logarithmic Negativity($LN$) and Reyni-entropy($S_{\delta}$). We present some useful bounds here. \\

\textbf{Theorem 3}: \emph{$LN(|\Gamma\rangle_{AB})\geq \frac{1}{2}\{LN(|\psi\rangle_{AB})+LN(|\phi\rangle_{AB})\}+2+\log\alpha\beta$}.\\

\textbf{Theorem 4}: \emph{$2\log(3(\alpha+\beta)(\min(\xi)))\leq LN(|\Gamma\rangle_{AB})\leq 2\log(3(\alpha+\beta)(\max(\xi)))$}, where $\min(\xi)=\min\{\sqrt{a_{i}},\sqrt{b_{i}}\}_{i=0}^{2}$ and $\max(\xi)=\max\{\sqrt{a_{i}},\sqrt{b_{i}}\}_{i=0}^{2}$.\\

\textbf{Theorem 5}: \emph{$S_{\delta}(|\Gamma\rangle_{AB})\geq \frac{\ln\{3(\alpha\beta)^{2\delta}\}}{1-\delta}+S_{\delta}(|\psi\rangle_{AB})+S_{\delta}(|\phi\rangle_{AB})$}.\\

\textbf{Theorem 6}: \emph{$(\frac{2\delta}{1-\delta})\ln(\min(\eta))\leq S_{\delta}(|\Gamma\rangle_{AB})\leq (\frac{2\delta}{1-\delta})\ln(\max(\eta))$}, where $\min(\eta)=\min\{\alpha\sqrt{a_{i}}+\beta\sqrt{b_{i}}\}_{i=0}^{2}$ and $\max(\eta)=\max\{\alpha\sqrt{a_{i}}+\beta\sqrt{b_{i}}\}_{i=0}^{2}$.\\

In\cite{subspace} Gour \textit{et. al.} derived bounds on the entanglement of the superposed state as a function of the entanglement of the components and von Neumann entropy($E$) of the reduced state of either party is taken as the measure of entanglement. From their work, we find the following upper and lower bounds.\\

\textbf{Theorem 7}: \emph{$E(|\Gamma\rangle_{AB})\leq(\alpha\sqrt{E(|\psi\rangle_{AB})+1}+\beta\sqrt{E(|\phi\rangle_{AB})+1})^{2}$}, with $E(|\psi\rangle_{AB})=S(tr_{A}(|\psi\rangle_{AB}\langle\psi|))=S(tr_{B}(|\psi\rangle_{AB}\langle\psi|)).$\\

In this context we have also found some upper bounds in two different forms; one is as a function of entanglement and other has a direct relation with the Schmidt coefficients of the states. \\

\textbf{Theorem 8}: \emph{$E(|\Gamma\rangle_{AB})+\alpha\log_{2}\alpha+\beta\log_{2}\beta\leq \alpha E(|\psi\rangle_{AB})+\beta E(|\phi\rangle_{AB})$}.\\

\textbf{Theorem 9}: \emph{$ E(|\Gamma\rangle_{AB})\leq 2[\log_{2}3(\alpha+\beta)]\max(\gamma)$}, where $\max(\gamma)=\max\{\sqrt{a_{i}},\sqrt{b_{i}}\}_{i=0}^{2}$.\\

We have already used the pairs ($|\psi\rangle_{AB}$, $|\psi'\rangle_{AB}$) and ($|\phi\rangle_{AB}$, $|\phi'\rangle_{AB}$)  in five different kinds of superpositions of comparable or incomparable states. From the above theorems we observe the upper and lower bounds of different entanglement measures and based on these observations we can form some counterintuitive situations of bounds which will be enough to establish the importance of the idea that comparability and incomparability under deterministic LOCC  that plays a crucial role in making the structure of the state space.\\

Let the pairs ($|\psi\rangle_{AB}$, $|\psi'\rangle_{AB}$) and ($|\phi\rangle_{AB}$, $|\phi'\rangle_{AB}$) have same entanglement, i.e., $E(|\psi\rangle_{AB})=E(|\psi'\rangle_{AB})$ and $E(|\phi\rangle_{AB})=E(|\phi'\rangle_{AB})$. This fact clearly indicates that both the pairs ($|\psi\rangle_{AB}$, $|\psi'\rangle_{AB}$) and ($|\phi\rangle_{AB}$, $|\phi'\rangle_{AB}$) are incomparable to each other (if we consider they have different Schmidt coefficients). We can construct infinitely many incomparable pairs of ($|\psi\rangle_{AB}$, $|\psi'\rangle_{AB}$) and ($|\phi\rangle_{AB}$, $|\phi'\rangle_{AB}$) with the above property. Specifically, for $\alpha=\alpha'$ and $\beta=\beta'$ Theorem 7 establishes the fact that both $|\Gamma\rangle_{AB}$ and $|\Gamma'\rangle_{AB}$ have the same upper bounds but at the same time Theorem 9 indicates another feature of the upper bounds. As the pairs ($|\psi\rangle_{AB}$, $|\psi'\rangle_{AB}$) and ($|\phi\rangle_{AB}$, $|\phi'\rangle_{AB}$) are incomparable to each other, we have either  $a_{0}>\alpha_{0}$ and $b_{0}>\beta_{0}$ or $a_{0}<\alpha_{0}$ and $b_{0}<\beta_{0}$. Therefore, for the first case we have upper bound of entanglement of ($|\Gamma\rangle_{AB}$) $\geq $ upper bound of entanglement of ($|\Gamma'\rangle_{AB}$) and for the later case upper bound of entanglement of ($|\Gamma\rangle_{AB}$)$\leq$ upper bound of entanglement of ($|\Gamma'\rangle_{AB}$).\\

Similar features could also be observed by considering the same entanglement of the pairs ($|\psi\rangle_{AB}$, $|\psi'\rangle_{AB}$)and ($|\phi\rangle_{AB}$, $|\phi'\rangle_{AB}$) and all other combinations of choice of the pairs ($|\psi\rangle_{AB}$, $|\psi'\rangle_{AB}$) and ($|\phi\rangle_{AB}$, $|\phi'\rangle_{AB}$) with respect to the idea of comparability and incomparability under deterministic LOCC for any arbitrary choice of $\alpha,\beta,\alpha'$ and $\beta'$.\\

Again we could employ the notion of incomparability from a different point of view to construct some new bounds of these different measures. Let the pairs ($|\psi\rangle_{AB}$, $|\psi'\rangle_{AB}$)and ($|\phi\rangle_{AB}$, $|\phi'\rangle_{AB}$) are incomparable to each other. So Negativity of the both pairs have the following relations:
$N(|\psi\rangle_{AB})\gtreqless N(|\psi'\rangle_{AB})$ and $N(|\phi\rangle_{AB})\gtreqless N(|\phi'\rangle_{AB})$. Now if we consider the following cases, i.e., $N(|\psi\rangle_{AB})\geq N(|\psi'\rangle_{AB})$ and $N(|\phi\rangle_{AB})\geq N(|\phi'\rangle_{AB})$ or $N(|\psi\rangle_{AB})\leq N(|\psi'\rangle_{AB})$ and $N(|\phi\rangle_{AB})\leq N(|\phi'\rangle_{AB})$, then using Theorem 1 we can find some tight upper and lower bounds of $N(|\Gamma\rangle_{AB})$ and $N(|\Gamma'\rangle_{AB})$ for any arbitrary choice of $\alpha,\beta,\alpha'$ and $\beta' $, assuming the other restrictions. In this case, when $\alpha=\alpha'$ and $\beta=\beta' $ we find the following relations\\
\begin{multline}
$$\frac{1}{2}[9(\alpha+\beta)^{2}\{min(\alpha_{2},\beta_{2})^{2}\}-1]\leq \frac{1}{2}[9(\alpha+\beta)^{2}\{min(a_{2},b_{2})^{2}\}-1]\leq \{N(|\Gamma'\rangle_{AB}) or N(|\Gamma\rangle_{AB})\}\\\leq \{N(|\Gamma\rangle_{AB}) or N(|\Gamma'\rangle_{AB})\}\leq \frac{1}{2}[9(\alpha+\beta)^{2}\{max(\alpha_{0},\beta_{0})^{2}\}-1]\leq \frac{1}{2}[9(\alpha+\beta)^{2}\{min(a_{0},b_{0})^{2}\}-1]$$.\\
\end{multline}
Similar types of features for the bounds could be observed for Logarithmic Negativity and Reyni's Entropy employing the above theorems and the comparability and incomparability relations with arbitrary choice of $\alpha,\beta,\alpha'$ and $\beta' $.
\section{CONCLUSION}

In conclusion we have observed that superposition of states may lead to pairs of incomparable states to a pair of comparable states under deterministic LOCC. Therefore, through the superposition of states we have succeeded in making a connection between two classes of states, i.e., comparable and incomparable. This technique would be useful in many aspects where we have some definite kind of tasks with some states which are incomparable in nature, however we could find a new pair that are comparable in nature through superposition. Since, incomparability may be used as a detection of un-physical operations \cite{general}, therefore, through the superposition we could form new classes of incomparable states to act as detector of un-physical operations. Also some one could find tighter bounds on entanglement behaviour through the superposition and using different comparable or incomparable pair of states.

{\bf Acknowledgement.} The author A. Sen acknowledges the financial support from University Grants Commission, New Delhi, India.


\begin{thebibliography}{99}
\bibitem{rosen} A.Einstein, B.Podolsky and N.Rosen,``Can a quantum-mechanical description of physical reality be considered complete?", {\it Physical Review A}, 47,777-780(1935).
\bibitem{sch} E.$Schr\ddot{o}dinger$, ``Die Gegenwartige situation in der qunteunme chnik ", {\it $Nat\ddot{u}rwissenschaften$}, Vol. 23, 807 (1935), doi.10.1007/BF0149189.
\bibitem{bell} J.S.Bell, ``On the Einstein Podolsky Rosen Paradox", {\it Physics}, Vol. 1, 195-200 (1964).
\bibitem{zurek} W. H. Zurek, ``Quantum cloning : $Schr\ddot{o}dinger's$ sheep", {\it Nature}, Vol.404, 130-131 (2000), doi.10.1038/35004684 .
\bibitem{nielsen} M. A. Nielsen, ``Condition for a class of entanglement transformations", {\it Physical Review Letter}, Vol.83, 436-439 (1999), doi.10.1103.
\bibitem{tele} Edited by D.Bouwmeester, A.Ekert and A.Zeilinger, {\it The physics of Quantum Information: Quantum Cryptography,Quantum Teleportation and Quantum Computation}, Springer, New York, (2000).
\bibitem{horo} M. Horodecki, R. Horodecki, A. Sen(De) and U. Sen, ``Common Origin of No-Cloning and No-Deleting Principles Conservation of Information", {\it Found. Phys.}, Vol. 35, 2041-2049 (2005).
\bibitem{Enk} S. J. van Enk, ``Relation between cloning and the universal NOT derived form conservation laws", {\it Physical Review Letter}, Vol. 95,  010502(2005), doi. 10.1103/Phys Rev Lett.95.010502.
\bibitem{brus} D. Bru$\beta$, D. P. DiVincenzo, A. Ekert, C. A. Fuchs, C. Macchiavello and J. A. Smolin, ``Optimal universal and state depandent quantum cloning", {\it Physical Review A}, Vol.57, 2368-2378 (1998), doi.10.1103/Phys Rev A.57.2368.
\bibitem{linear} D. Dieks, ``Communication by EPR devices",  {\it Physics Letter A} ,Vol. 92, 271-272 (1982), doi.0.031-9163/82/0000-0000/02.75.
\bibitem{super} N.Linden, S.Popescu and J.Smolin, ``Entanglement of superposition", {\it Physical Review. Letter}, Vol.97, 100502(2000), doi.10.1103/Phys. Rev. Lett. 97, 100502.
\bibitem{unitary} H. P. Yuen, ``Amplification of quantum states and noiseless photon amplifiers", {\it Physics Letter A} , Vol. 113, 405-407 (1986), doi.10.1016/0375-9601(88)90660-2.
\bibitem{gisin} N. Gisin and S. Popescu, ``Spin flips andquantum information foe antiparallel spins",  {\it Physical Review Letter}, Vol.83, 432-435 (1999), doi.10.1103/Phys Rev Lett.83.432.
 \bibitem{flip} I. Chattopadhyay, S. K. Choudhary, G. Kar, S. Kunkri and D.Sarkar, ``No fliping as a consequence of no-signalling and non-incresement of entanglement under LOCC", {\it Physics Letter A}, Vol.351, 384-387 (2005), doi.10.1016/J.Physleta.2005.11030.
\bibitem{incompare} I. Chattopadhyay and D. Sarkar, ``Deterministic local conversion of incomparable states by collective LOCC", {\it Quantum Information and Computation}, Vol.5,  247-257 (2005), doi.10.1007/S11128-008-0085-6., available online in arXiv:quant-ph/0409174.
\bibitem{incomfi} I. Chattopadhyay and D. Sarkar, ``Impossible of exact flipping of three arbitrary quantum states via incomparability", {\it Physical Review A}, Vol.73, 044303(2006), doi.10.1103/arXiv:quant-ph/0511040.
\bibitem{nielsen1} M. A. Nielsen and I. L. Chuang, {\it Quantum Computation and Quantum Information}, Cambridge University Press, (2000).
\bibitem{bound1}Chang-shui Yu, X. X. Yi, He-shan Song, "Bounds on bipartitely shared entanglement reduced from superposed tripartite quantum states ", {\it Euro. Phys. J. D}, Vol.49,273-278, DOI: 10.1140/epjd/e2008-00162-7 .
 \bibitem{con}J. Niset and N. J. Cerf, "Tight bounds on the concurrence of quantum superpositions", {\it Physical Review A }, Vol.76, 042328 (2007).
 \bibitem{con1}Chang-shui Yu, X. X. Yi and He-shan Song, "Concurrence of superpositions", {\it Physical Review A }, Vol.75, 022332 (2007).
 \bibitem{negativity}Yong-Cheng Ou and Heng Fan, "Bounds on negativity of superpositions", {\it Physical Review A }, Vol.76, 022320 (2007).
 \bibitem{multi}D. Cavalcanti, M. O. Terra Cunha, A. Acin, "Multipartite entanglement of superpositions", {\it Physical Review A }, Vol.76, 042329 (2007).
 \bibitem{multi1}Wei Song, Nai-Le Liu and Zeng-Bing Chen, "Bounds on the multipartite entanglement of superpositions", {\it Physical Review A }, Vol.76, 054303 (2007).
 \bibitem{ent}Gilad Gour, "Reexamination of entanglement of superpositions", {\it Physical Review A }, Vol.76, 052320 (2007).
 \bibitem{subspace}Gilad Gour and Aidan Roy, "Entanglement of subspaces in terms of entanglement of superpositions", {\it Physical Review A }, Vol.77, 012336 (2008).
\bibitem{som} S. Bandyopadhyay, V. Roychowdhury and  U. Sen, {\it Physical Review A}, Vol.65, 052315 (2002).
\bibitem{general} I. Chattopadhyay and D. Sarkar, ``General Classes of Impossible Operations
through the Existence of Incomparable states", {\it Quantum Information and Computation}, Vol.6,  93-99 (2007).
\end{thebibliography}
\end{document}